A Comprehensive Analysis on Hurricane Lane 2018's Peak Intensity Using Recon Data, Satellite, and

Microwave Imagery

Michael A. Igbinoba

August 3, 2021

Table of Contents

**Abstract**

The recon data of Hurricane Lane, similar to the likes of several other systems like Hurricane Dorian of 2019, Hurricane Felix of 2007, and Hurricane Matthew of 2016, shows a large discrepancy between flight level wind data and SFMRs. This large discrepancy has caused uncertainty over Lane's peak intensity. In an attempt to capture Lane's peak intensity, the author of this analysis has created a vertical wind profile using dropsondes that would suggest a flight level to surface conversion of x0.97, higher than the original x0.9 from Franklin's original study. The author has also analyzed flight level wind data, SFMR data, other dropsonde data, microwave imagery, and satellite imagery for results that would support a 145kt MSW and 926hPa MSLP. The results of this analysis can be used to determine the true reliability of instruments like SFMR and dropsondes in a comprehensive analysis. Furthermore, the results of this analysis can be used to determine potential weak points in a tropical cyclone's structure that could cause recon to pick up lower values than in other areas despite slow movement, and how abrupt changes in satellite appearance can be used to infer a more or less dramatic weakening cycle, via satellite imagery analysis. Finally, the results of this analysis can be used to infer how certain features a tropical cyclone exhibits can affect recon data output.

**Peak Recon Data, and Microwave Analysis**

<u>Vertical Wind Profile Analysis</u>

During the peak mission, recon flew at the 740-760hPa level. Dropsonde vertical wind profile research (Franklin et al. 2002) allowed an empirical flight level to surface profile to be created from Guillermo and Erika in 1997, Bonnie, Danielle, Georges, Mitch, Lester, and Madeline in 1998, and Bret, Dennis, Floyd, Gert, Irene, Jose, Lenny, Dora, and Eugene in 1999. Furthermore, this allowed for the creation of standardized conversions for the 700hPa, 850hPa, 925hPa, and 1000ft flight levels that are used operationally by the National Hurricane Center. This is all shown below:

| Flight level | Eyewall | Outer vortex (convection) | Outer vortex (not in convection) |
|---|---|---|---|
| 700 hPa | 0.90 | 0.85 | 0.80 |
| 850 hPa | 0.80 | 0.80 | 0.75 |
| 925 hPa | 0.75 | 0.75 | 0.75 |
| 1000 ft (305 m) | 0.80 | 0.80 | 0.80 |

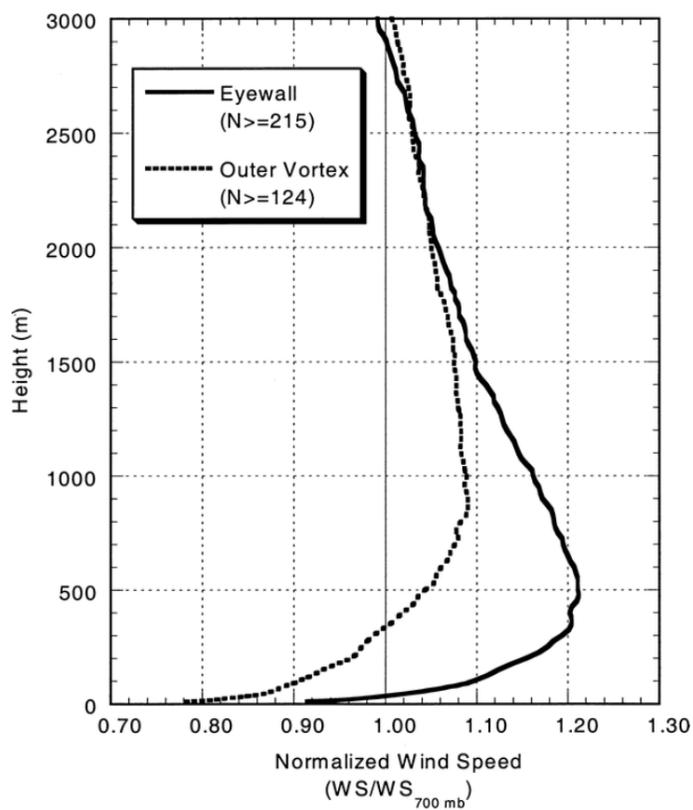

*Figures 1 & 2: The accepted standardized flight level/surface conversion factors also used operationally by the National Hurricane Center, and mean vertical wind profile resulting from the vertical wind profiles of 17 recon-observed tropical cyclones.*

Due to winds at the 740hPa level having no standardized conversion, the mean wind profile must be used. Using a factor of 0.91 for an eyewall observed wind, divided by 1.05 on the eyewall trend line for a height of ~2280 meters (0.91/1.05) achieves a 740hPa to surface conversion of x0.86. However, with accurate and a reasonable amount of dropsondes available throughout Lane's observed period (August 21 & 22 2100-1000 UTC), and the mean profile's empirical nature, the author has replicated the study and made his own vertical wind profile.

The vertical wind profile for Lane was constructed with 7 eyewall dropsonde containing 210 datapoints, across missions 9, 11 & 12. All dropsondes met the criteria of recording a surface minimum pressure ≤ 960hPa, all gusts ≥ 100kt, and no influence from transient features such as mesovortices. All data was obtained from Tropical Atlantic's recon archive. The constructed profile is shown on the next page:

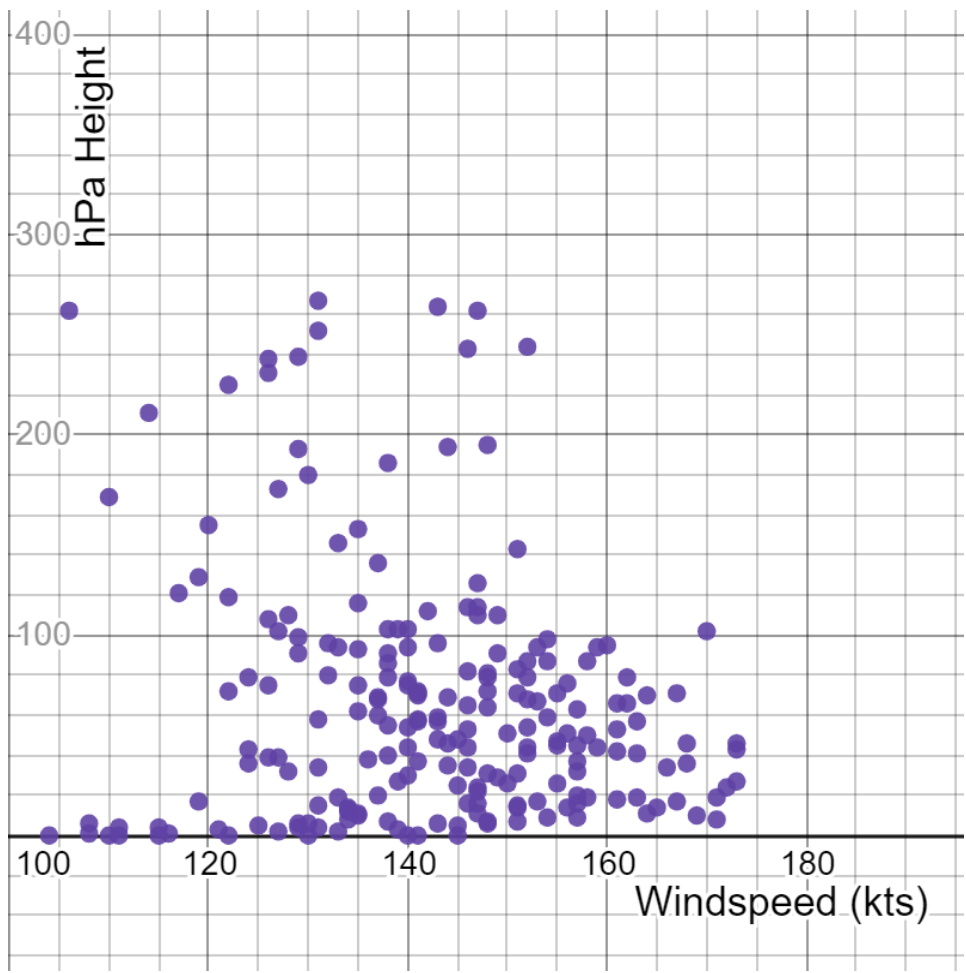

*Figure 3: A constructed vertical wind profile consisting of 210 datapoints from 9 eyewall dropsondes, all meeting requirements.*

Something that can be noticed with the dataset, is the boundary layer top height, shown using a trend line below:

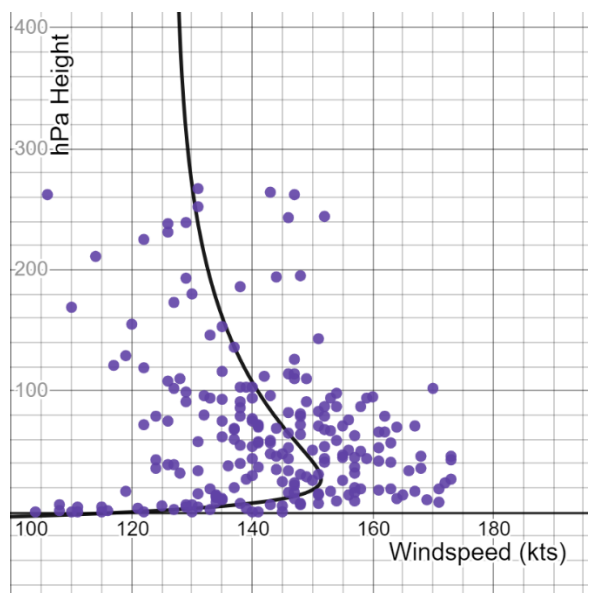

*Figure 4: A trend line plotted which averages the profile data on a curve. The line was constructed using an x=ab^y-cd^y+f equation on the Desmos regression calculator.*

Using the data points and trend suggests a maximum boundary layer height of ~40hPa above surface, or ~400 geopotential meters. Boundary layer maximum height is usually found to be 50hPa or 500 geopotential meters. However, nothing else seems to stand out about the data so far.

The gust data will now be set against the maximum 740-760hPa flight level winds at the time of each drop to achieve a multiplier, shown on the next page on the thoroughly smoothed profile:

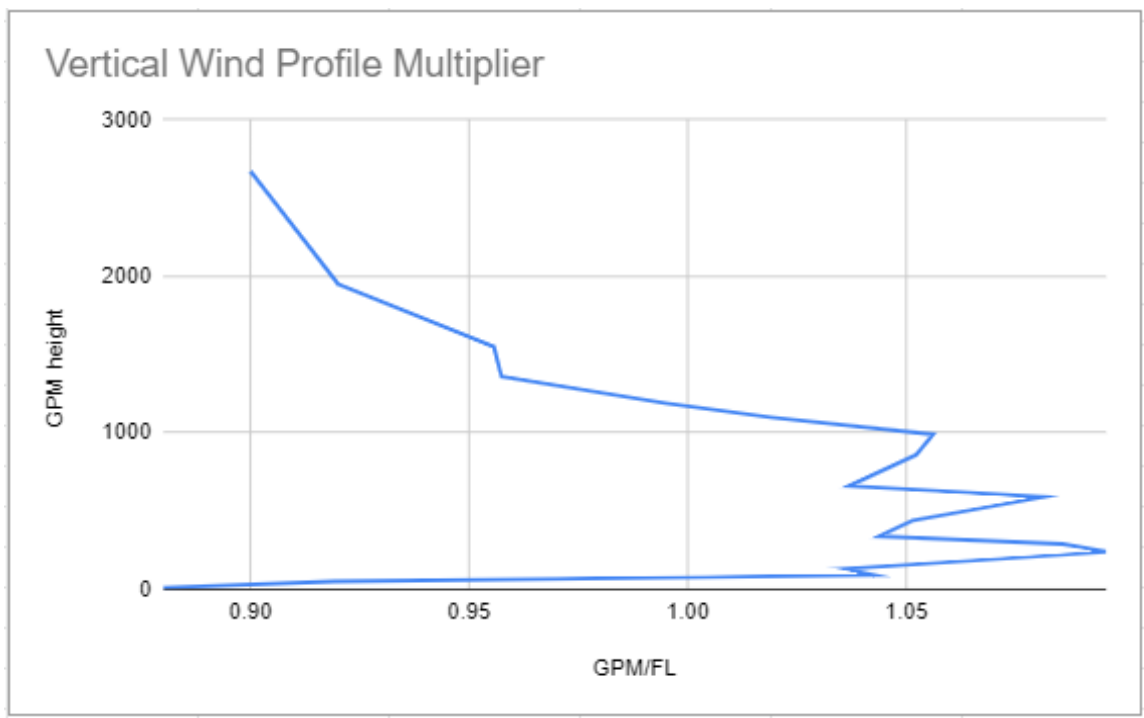

*Figure 5: Dropsonde gusts set against 740-760hPa winds with its trend.*

There is something noticeable about the dataset again now that it is set against the 740hPa winds. Winds at the top of the boundary layer are ~1.02 times stronger than the winds at the 740hPa level. According to Franklin's study, this is unexpected. Winds in Franklin's study have been shown on several profiles to be weaker than what Lane exhibited. Unfortunately, this phenomenon goes without explanation.

Something else noticeable, are the jagged two additional peaks shown on the profile. These peaks could be the result of data noise, as some data could not be smoothed out to fix it.

To figure out the new 740hPa to surface conversion, the trend needs to be traced to its GPM at the time (2280 GPM) of the peak pass. And afterwards, the traced multiplier will divide the surface multiplier of 0.88.

The final conversion value achieved from the vertical wind profile, after dividing 0.88 by the 0.91 derived from the trend (0.88/0.91), is x0.97 for the 740hPa level, after considering that a slight majority (6/9) of the dropsondes were inbound, which would lower possibilities of the conversion ratio being skewed downward or upward. There could be a varying uncertainty range with the still slightly sparse dataset, but that will be addressed later in the analysis.

## Flight Level Data Analysis

During Lane's peak at around 0400 UTC, recon found strong flight level winds at 740-760hPa, in the northwest and eastern portions of the eyewall. The flight level winds that will be presented below and on the next page are from the Atlantic Oceanographic & Meteorological Laboratory's Hurricane Research Division:

| TIME   | Lat    | Lon     | Head  | Track  | GnSpd | TAS   | GeoAl  | Press | WndDr  | WndSp | Tempr | Dewpt | D Val  | RdAlt  | MixR    | VtWnd | SfcPr | ThetaE |
|--------|--------|---------|-------|--------|-------|-------|--------|-------|--------|-------|-------|-------|--------|--------|---------|-------|-------|--------|
| HHMMSS | Deg N  | Deg W   | Deg   | Deg    | m/s   | m/s   | m      | mb    | Deg    | m/s   | C     | C     | m      | m      | g/m3    | m/s   | mB    | Deg    |
| 035103 | 14.535 | 154.093 | 75.90 | 106.30 | 90.6  | 133.5 | 2262.0 | 742.8 | 34.60  | 73.70 | 13.12 | 14.70 | -280.0 | 2266.0 | 14.3700 | 4.90  | 964.4 | 356.00 |
| 035104 | 14.535 | 154.092 | 74.70 | 106.20 | 90.4  | 131.7 | 2264.0 | 741.9 | 34.60  | 71.40 | 13.52 | 14.80 | -287.1 | 2269.0 | 14.4600 | 4.60  | 963.2 | 356.90 |
| 035105 | 14.533 | 154.092 | 75.10 | 106.40 | 89.7  | 132.0 | 2268.0 | 741.6 | 32.90  | 74.60 | 13.45 | 14.90 | -287.2 | 2273.0 | 14.5500 | 4.10  | 963.2 | 357.10 |
| 035106 | 14.533 | 154.092 | 75.60 | 106.90 | 89.4  | 131.6 | 2272.0 | 741.0 | 32.20  | 75.80 | 13.45 | 15.00 | -289.8 | 2276.0 | 14.6500 | 4.40  | 962.8 | 357.50 |
| 035107 | 14.533 | 154.090 | 75.10 | 107.60 | 89.2  | 130.2 | 2276.0 | 740.3 | 31.60  | 75.60 | 13.59 | 15.00 | -292.4 | 2281.0 | 14.7000 | 6.60  | 962.3 | 357.90 |
| 035108 | 14.533 | 154.090 | 74.60 | 108.30 | 89.0  | 133.2 | 2282.0 | 739.4 | 33.60  | 78.30 | 13.06 | 15.10 | -296.9 | 2287.0 | 14.7700 | 7.20  | 962.1 | 357.60 |
| 035109 | 14.533 | 154.088 | 74.70 | 108.70 | 88.5  | 135.1 | 2289.0 | 738.4 | 36.40  | 78.60 | 12.99 | 15.10 | -300.8 | 2293.0 | 14.8200 | 6.50  | 961.6 | 357.80 |
| 035110 | 14.533 | 154.088 | 76.10 | 108.90 | 87.9  | 131.5 | 2295.0 | 738.0 | 35.70  | 75.70 | 13.40 | 15.00 | -299.2 | 2299.0 | 14.7800 | 9.30  | 961.4 | 358.30 |
| 035111 | 14.533 | 154.087 | 77.70 | 109.40 | 87.3  | 128.4 | 2303.0 | 736.8 | 33.10  | 75.50 | 13.40 | 15.00 | -303.2 | 2307.0 | 14.8000 | 5.40  | 960.8 | 358.50 |
| 035112 | 14.532 | 154.087 | 77.50 | 109.80 | 87.0  | 129.1 | 2309.0 | 736.3 | 37.80  | 72.10 | 13.29 | 15.00 | -302.5 | 2314.0 | 14.7900 | 8.50  | 960.9 | 358.40 |
|        |        |         |       |        |       |       |        |       |        |       |       |       |        |        |         |       |       |        |
| 035544 | 14.443 | 153.813 | 120.50| 86.10  | 111.7 | 130.5 | 2030.0 | 758.4 | 179.40 | 74.30 | 16.58 | 15.50 | -346.9 | 2034.0 | 14.8100 | -4.80 | 956.5 | 359.60 |
| 035545 | 14.443 | 153.812 | 121.00| 85.90  | 110.7 | 128.3 | 2031.0 | 758.5 | 180.50 | 74.20 | 17.13 | 15.70 | -344.7 | 2035.0 | 14.9700 | -4.40 | 956.3 | 360.80 |
| 035546 | 14.443 | 153.812 | 122.40| 85.60  | 109.8 | 128.3 | 2032.0 | 758.7 | 179.40 | 73.60 | 17.45 | 15.60 | -340.8 | 2036.0 | 14.9400 | -5.10 | 956.5 | 361.10 |
| 035547 | 14.443 | 153.810 | 122.80| 85.60  | 108.8 | 127.3 | 2033.0 | 759.0 | 179.90 | 74.60 | 17.77 | 15.40 | -337.3 | 2037.0 | 14.6800 | -3.90 | 956.7 | 360.70 |
| 035548 | 14.443 | 153.808 | 122.60| 85.60  | 107.8 | 126.4 | 2035.0 | 759.0 | 180.20 | 75.00 | 17.92 | 14.90 | -335.2 | 2039.0 | 14.2600 | -4.30 | 956.9 | 359.50 |
| 035549 | 14.443 | 153.808 | 122.50| 85.70  | 107.0 | 123.8 | 2036.0 | 759.1 | 181.80 | 74.30 | 17.99 | 14.60 | -333.2 | 2040.0 | 13.9600 | -3.30 | 957.1 | 358.70 |
| 035550 | 14.443 | 153.807 | 123.20| 85.90  | 106.5 | 122.9 | 2037.0 | 759.2 | 181.70 | 72.90 | 17.67 | 14.40 | -330.9 | 2041.0 | 13.7500 | -0.60 | 957.6 | 357.60 |
| 035551 | 14.443 | 153.807 | 123.70| 86.20  | 106.0 | 123.5 | 2039.0 | 759.2 | 181.20 | 73.20 | 17.20 | 14.50 | -329.1 | 2043.0 | 13.8100 | -1.20 | 958.2 | 357.20 |
| 035552 | 14.443 | 153.805 | 123.50| 86.50  | 105.6 | 124.5 | 2042.0 | 759.4 | 180.90 | 74.50 | 17.05 | 14.70 | -324.4 | 2046.0 | 14.0500 | -1.60 | 958.8 | 357.70 |
| 035553 | 14.443 | 153.803 | 123.40| 86.60  | 105.4 | 120.8 | 2044.0 | 759.8 | 184.00 | 73.90 | 17.61 | 15.10 | -317.3 | 2048.0 | 14.3800 | -3.10 | 959.1 | 359.40 |

*Figure 6 & 7: Flight level winds during 0351 UTC in the northwest portion of the eyewall, and flight level winds during 0355 UTC in the eastern portion of the eyewall.*

At 035109 UTC in the northwest portion of the eyewall, recon recorded a maximum wind of 78.6 meters per second, or 152.8kts. However, this is one second data. For winds to be countable as representative of a storm's 1-min intensity, a 10 second average must be taken.

Averaging the values in the dataset achieves a 10 second average of 75.1 meters per second, or 146.0kts. Taking an average of the radar altitude values gets a mean height of 2282 meters. Using this height on the vertical wind profile gets a surface conversion of x0.97. Applying the conversion to the 146kt 10 second average gets a surface value of 141.6kts. Without considering anything outside of the flight level data and vertical wind profile results, this would support a 140kt intensity at Lane's peak.

At 035548 UTC in the eastern portion of the eyewall, recon recorded a maximum wind of 75.0 meters per second, or 145.8kts. However, once again this is one second data. A 10 second average must be taken from the available data. Averaging the values in the dataset gets a 10 second average of 74.0 meters per second, or 143.9kts. Taking an average of the radar altitude values gets a mean height of 2036 meters. Using this height on the vertical wind profile gets a surface conversion of x0.96. Applying the conversion to the 143.9kt 10 second average gets a surface value of 138.1kts.

Something that stands out upon further inspection of flight level data, is how axisymmetric Lane's wind structure seems to be. Based on the previous averaging of flight level winds in both the northwest and east eyewall sections, there is a small difference of 2.1kts between them, with the northwestern section being the stronger section. This would also make sense given the storm's movement speed at 6kts due west during the time coupled with the winds not being directly to the north of the center, but northwest which should lower the kinetic boost such winds receive. Furthermore, this symmetry is also shown when plotting the flight level winds, as shown below:

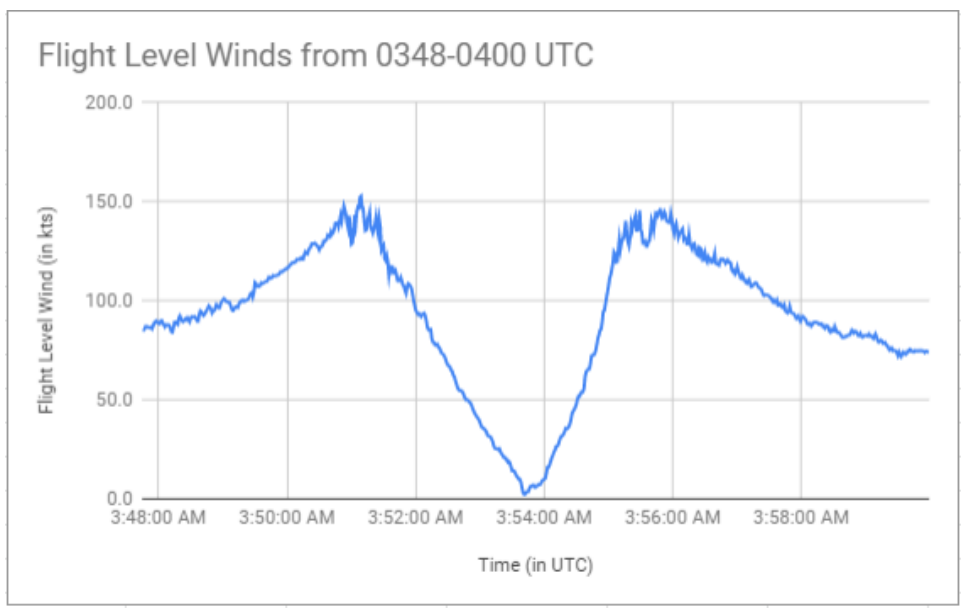

*Figure 8: Plotted flight level winds showing a near-axisymmetric wind structure in the eyewall.*

Upon further inspection of the plot, there seems to be two well-defined deformities in winds for both sections. This is due to recon suddenly hitting a stronger patch of updraft with velocities above 10-15 meters per second, which is exceptionally powerful to the point where it would cause turbulence that can briefly offset flight level winds.

When converting the peak flight level winds using the vertical wind profile, there is also a 3.5kt difference between the northwest and east section. This would still show a symmetrical eyewall. However, it is not a direct measurement like flight level winds, so less emphasis can be placed on such. This will be referred to again later in the analysis.

Returning to a point mentioned earlier in the analysis, there are some inherent issues with having a smaller dataset than what would be considered optimal. Therefore, a confidence interval must be determined. Using a 0.98/sqrt (dataset size) equation for 210 datapoints, yields a 95% confidence interval of 6.76%. This would apply to the 141.6kts and 138.1kts surface winds resulting from the vertical wind profile conversion multiplier, and would not apply to the conversion multiplier itself. Confidence rates and conversion results are in the following table:

| Confidence Rates & Conversion Results | "Low" 95% | Initial Conversion | "High" 95% |
|---|---|---|---|
| Northwest Section | 132.0kts | 141.6kts | 151.2kts |
| East Section | 128.8kts | 138.1kts | 147.4kts |

*Figure 9: A chart showing low- and high-end confidence rates for each recorded section of the eyewall, resulting from a less than optimal amount of datapoints.*

## SFMR Data Analysis

During Lane's peak around 0400 UTC, SFMR began to record extreme values far stronger than the maximum flight level winds themselves and what they suggested at surface, instigating a cause for investigation. The data that will be displayed below is once again from the Atlantic Oceanic Meteorological Laboratory's Hurricane Reanalysis Division:

```
Time    RR      WS(m/s) | kts)   TB1     TB2     TB3     TB4     TB5     TB6
35129   14.5    78.4      152.4   212.9   221.3   224.7   231.9   238.5   244.2
35130   11.8    78.8      153.2   214.1   221.3   224.7   231.9   238.5   243.4
35131   9.3     79.2      154.0   214.1   221.9   227     231.9   238.5   243.4
35133   7.6     79.3      154.1   214.1   221.9   227     231.5   237.6   243.4
35134   5.5     79.5      154.5   213.4   221.9   227     231.5   237.6   240
35135   3.6     79.6      154.7   213.4   220.9   224.9   231.5   237.6   240
35136   2.3     79.6      154.7   213.4   220.9   224.9   230.3   235.1   240
35137   1.3     79.5      154.5   212.5   220.9   224.9   230.3   235.1   238.3
35138   0.6     79.3      154.1   212.5   219.1   222.1   230.3   235.1   238.3
35139   0.6     79        153.6   212.5   219.1   222.1   225.9   230.7   238.3

35529   21.9    65.5      127.3   196.2   202.8   206.9   212.9   219.3   227.1
35530   21.5    65.6      127.5   196.7   202.8   206.9   212.9   219.3   225.2
35531   21.3    65.8      127.9   196.7   203.2   207.7   212.9   219.3   225.2
35532   21.2    65.9      128.1   196.7   203.2   207.7   213.1   221.5   225.2
35533   20.9    66.1      128.5   197.1   203.2   207.7   213.1   221.5   227.3
35534   21.1    66.1      128.5   197.1   204.8   209.2   213.1   221.5   227.3
35535   21.7    66.1      128.5   197.1   204.8   209.2   215.2   222.9   227.3
35536   22.5    66        128.3   198.2   204.8   209.2   215.2   222.9   229
35537   23.4    65.9      128.1   198.2   205.8   208.9   215.2   222.9   229
35538   24.2    65.7      127.7   198.2   205.8   208.9   214.5   223.1   229
35539   25      65.5      127.3   197.8   205.8   208.9   214.5   223.1   229.8
```

*Figure 10 & 11: SFMR winds during 0351 UTC in the northwest portion of the eyewall, and SFMR winds during 0355 UTC in the eastern portion of the eyewall.*

At 035135 UTC in the northwest section of the eyewall, SFMR estimated 154.7kt winds for two seconds, which marks a clear discrepancy between flight level winds and their conversions, and SFMR winds. A 10 second average of the available data gets a 10 second wind of 153.8kts, which would still be remarkably higher than flight level winds and their conversions.

Something that is noteworthy about the data, is that there seems to be little rain affecting it, with a maximum rain rate of 3.6 mm/hr. during the peak values, lowering the risk of direct contamination by higher rain rates. Without further inspection, these values would be considered accurate to Lane's intensity.

However, at 035533 UTC in the east section of the eyewall, SFMR estimated 128.5kt winds for three seconds, which is much lower in comparison to the flight level winds recorded at the time. A 10 second average of the available data gets a 10 second wind of 128.0kts. This data has a high rain rate at peak winds of 21.7 mm/hr. This rain rate is potentially enough to cause issues with the recorded data, thus making it less reliable without further review.

Something that immediately makes this data suspect, is the large asymmetry in winds indicated by it, shown below:

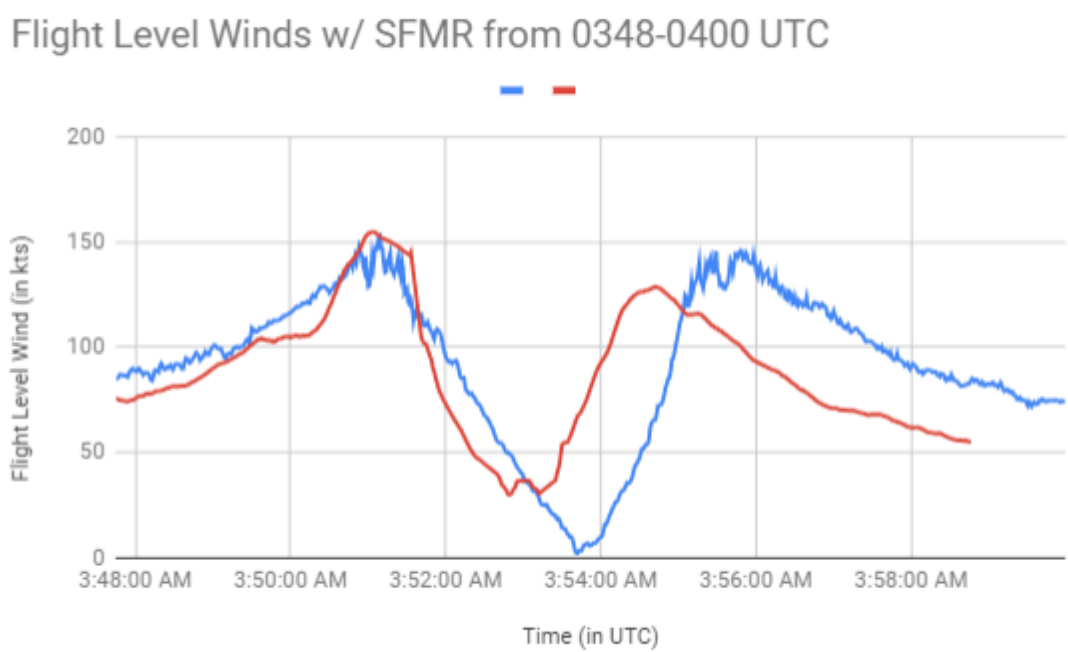

*Figure 12: SFMR winds with flight level winds, indicative of a large asymmetry between the sections.*

SFMR estimates compared to the flight level winds seem to be far less symmetrical, peaking at 154.7kts in the northwest section, while peaking at 128.5kts in the east section, while flight level winds peak at 152.8kts in the northwest section, and 145.8kts in the east section. Wind structure at the surface should still be symmetrical similarly to flight level winds, as higher precipitation indicated by SFMR should cause an easier mixdown of the same symmetrical flight level winds (also shown by the previous vertical wind profile) and the core structure's symmetry also promotes a higher level of symmetry than shown by SFMRs, but this will be talked about further later in the analysis. Quadrant wind boosts from kinetic processes as the storm moves at 6kts would not make such a difference either, as already shown by flight level winds. Regardless, lack of symmetry compared to actual flight level measurements is part of what makes the data suspect.

One additional thing that can be noticed with the plot above, is that SFMR seems to peak later than flight level wind in the northwest section, and before the flight level wind in the east section. This is because Lane exhibited an extreme stadium effect, visible on satellite imagery, and to reconnaissance aircraft on Vortex Data Messages, shown below:

**Remarks Section - Additional Remarks…**

**PENETRATION AT 8000 FT**
**STADIUM EFFECT IN EYE**

*Figure 13: Mission 11 Vortex Data Messages highlighting a stadium effect in the eye at Lane's peak.*

A stadium effect would cause surface and flight level winds to peak at contrasting times, with surface winds being amplified by this effect, as stadium effects curve inward towards the hurricane's center, which would cause higher altitudes to get into the storm's radius of maximum wind earlier than surface. This would cause stronger than normal surface winds because a smaller radius of maximum wind at surface compared to at flight level would allow for conservation of angular momentum to boost winds through acceleration as the radius of maximum winds contract to surface. This could also partially explain the much lower rain rates than expected, as a stadium effect would allow for less precipitation in its vicinity and as a result, cause less rain rates. One last thing that could have caused the lower rain rates, is the extreme updraft velocity sampled, which went above 15 meters per second at times. This could have cancelled out the downdraft motion caused by precipitation in the stadium, and caused less precipitation to occur.

Another thing that makes the data suspect is the relationship between the temperature brightness spikes, and rain rates, plotted below:

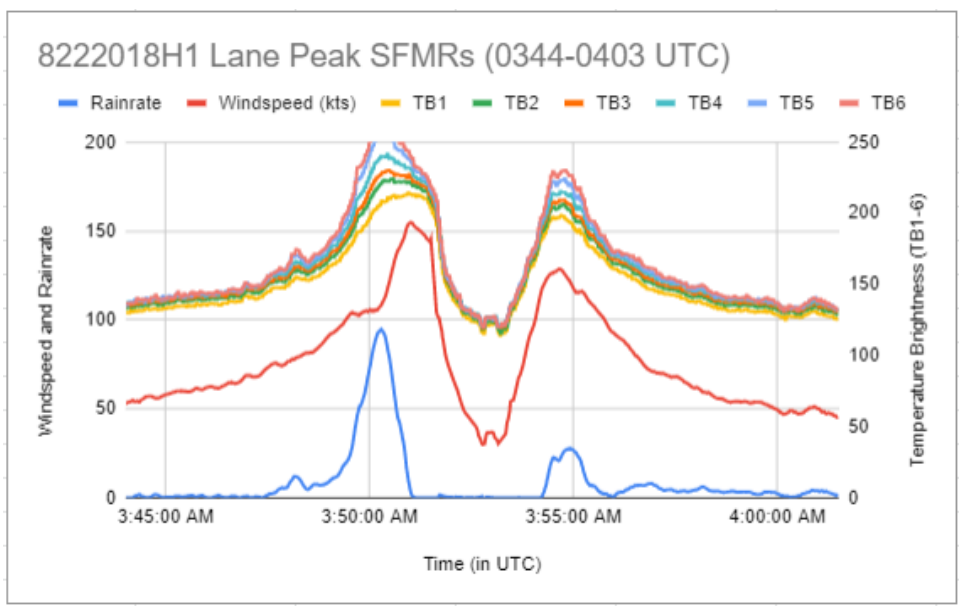

*Figure 14: SFMR wind data plotted with temperature brightness and rain rates.*

This plot is showing a distinct correlation between rain rate spikes and temperature brightness signal. SFMR measures brightness temperature from sea roughness, then processes it with its several channels, along with taking air temperature, salinity, altitude, and SST to produce a wind speed and rain rate estimate. However, rain is still a problem for the algorithm, as it can mistake precipitation for warmer brightness temperatures resulting from ocean roughness. This can cause an inflation of the wind estimate it produces, causing inaccuracies. The first major temperature brightness spike in the data set as it enters the eyewall, is closely correlated with an extreme spike in rain rates at 100mm/hr. Furthermore, this spike could have caused issues for the data that followed, specifically the 140kt+ SFMRs as it inflated temperature brightness values, which should also inflate wind estimates. SFMR wind estimates peaked as likely inflated brightness temperatures peaked, so it is unlikely that the algorithm accounted for this inflation, and this would aid in deeming the 150kt+ SFMRs less accurate to Lane's intensity.

In the east section of the eyewall, there is not as much of a rain spike as in the northwest section of the eyewall. As a result, there seems to be cooler temperature brightness signals with much less rain to inflate them. This helped result in the significantly weaker SFMR estimates of 128.5kts and the large asymmetry between the northwest section of the eyewall and east section.

In mid-2021, SFMR research aimed at trying to find a potential upward bias of SFMRs at higher intensities was successful, and found a varying upward bias associated with the 2014 algorithm update by using new sonde data, and creating a new wind-induced emissivity curve shown on the next page:

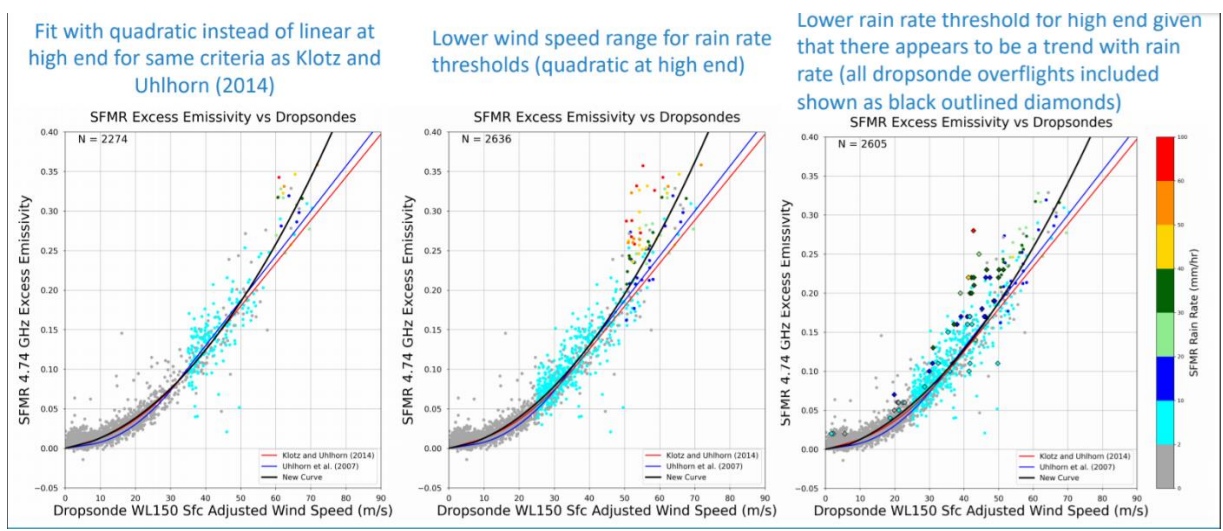

*Figure 15: New quadratic emissivity curve from recent SFMR research confirming an upward bias in the current algorithm (2014).*

Due to biases above the 120-130kt threshold when juxtaposed to new and old WL150s, it is also highly likely that Lane was on the receiving end of this upward bias, and this would reduce the reliability of the SFMRs even more. Overall, SFMRs for Lane's peak due to all the aforementioned reasons, should not be taken at face value, and should be used in junction with other reconnaissance data.

Dropsonde Analysis

There were two eyewall dropsondes at the time of Lane's peak. Unfortunately, due to the recording nature of sondes, where they must be positioned very well within the radius of maximum winds to get WL150s and MBL500s accurate to a system's intensity, and the time they were dropped, they recorded weaker winds that were unrepresentative of Lane's intensity, and therefore will not be used in this analysis outside of the previous vertical wind profile.

Although there are no useful eyewall dropsondes to estimate winds from, there is a center dropsonde from the peak mission that recorded the minimum pressure of Lane, shown below:

| Significant Wind Levels | | |
|---|---|---|
| Level | Wind Direction | Wind Speed |
| 927mb (Surface) | 165° (from the SSE) | 13 knots (15 mph) |
| 922mb | 190° (from the S) | **15 knots (17 mph)** |
| 850mb | 205° (from the SSW) | 13 knots (15 mph) |
| 817mb | 210° (from the SSW) | 10 knots (12 mph) |
| 796mb | 165° (from the SSE) | 10 knots (12 mph) |
| 754mb | 165° (from the SSE) | 6 knots (7 mph) |

*Figure 16: Dropsonde released in the eye of Lane at its peak, from mission 11, observation 7.*

The dropsonde recorded a pressure of 927hPa at surface. However, this is not the storm's true central pressure. The 927hPa pressure was accompanied by a 13kt gust. A 13kt gust would signify that this was not in the exact center of the cyclone, and that the pressure was lower than what the dropsonde recorded. A brief guide for adjusting central pressure based on gusts is shown on the next page:

```
0-10kt = 0mb
10-20kt = 1mb
20-30kt = 2mb
30-40kt = 3mb
40-50kt = 4mb
50-60kt = 5mb
etc...
```

*Figure 17: A guide on how many mb to adjust pressure down by based on recorded gust (Note: All pressures must be in the eye to work properly, as it is made for the even pressure gradience that occurs within the eye.)*

The conversion guide on the previous page would suggest a 1 hPa adjustment to the 927hPa recording. This would arrive at a central pressure of 926hPa at Lane's peak.

Microwave Imagery & Recon Data

Referring back to a statement made earlier in the analysis about Lane's inner core symmetry, microwave imagery passed from SSMI/S at 0339 UTC, shown below:

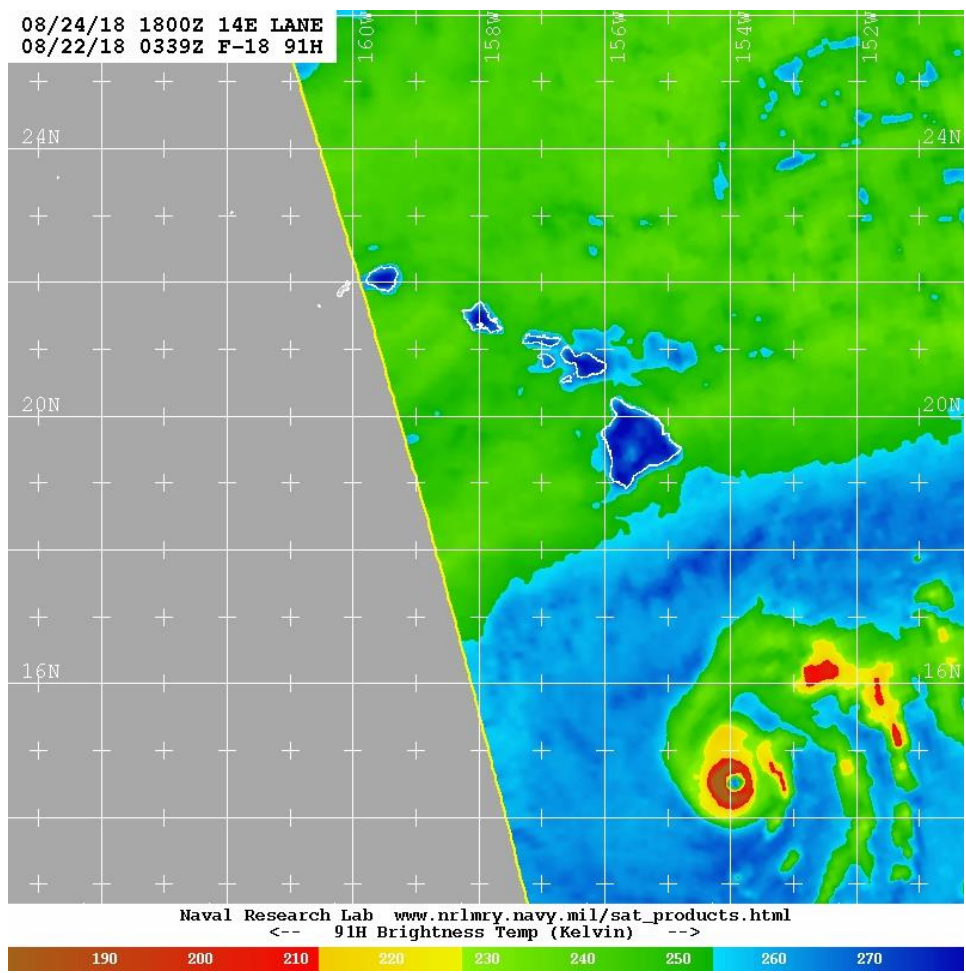

*Figure 18: SSMI/S pass at 0339 UTC revealing Lane's inner core.*

The above microwave image shows a compact, symmetric core structure which would be typical of systems that are Lane's strength, as their maturity allows for a more idealized lift process that allows little asymmetry in the inner core and CDO structure. This symmetry would further support the symmetric wind structure indicated by flight level winds and its conversions. However, there seems to be a slight bulge to the west. This could be ruled as something against its wind symmetry, as it immediately appears the northwest is much stronger. However, this is not due to storm-produced updraft asymmetries, but the westerly shear it was facing at the time. The westerly shear is apparent on the microwave imagery already with lack of any banding to the west, as any banding would naturally be suppressed. Another slight contributor of the lack of banding could be dry air to the north and west of the system, stabilizing the environment to its west and stymieing banding development. However, sea surface temperatures of 28-29C at the time would cause enough surface heat flux to generate banding to the west regardless of dry air. Therefore, it is most likely westerly shear. This westerly shear could have easily forced convective updrafts that would scatter ice far more on the west side, causing it to look far stronger when it is only slightly stronger than the east. Regardless, the storm's axisymmetric microwave conveys what the flight level winds convey, and is another valid argument to why SFMRs output is less trustworthy.

## Conclusion

The final products of the flight level data and SFMR data analyses will be used in a geometric mean equation, with one being weighted more than the other due to uncertainty bounds. Previously in the analysis, using a 95% confidence interval equation based on dataset size for the vertical wind profile arrived at a 6.76% confidence interval. SFMR has a far lower certainty than the vertical wind profile's 6.76% due to all the issues pointed out with how it operated in Lane. As a result, the SFMR data is objectively more uncertain while the vertical wind profile is more certain, and the winds resulting from the conversion achieved from the vertical wind profile will be double weighted in the equation shown below:

$$(141.6 \cdot 141.6 \cdot 153.8)^{\frac{1}{3}}$$

*Figure 19: The double weighted surface winds from flight level conversions and peak 10 second average of SFMR winds in a geometric mean equation.*

The results of the equation on the previous page yields a 145.6kt intensity for Lane's peak intensity, which rounds down to 145kts. The pressure of 927hPa adjusted downward by 1hPa to 926hPa will also be used in this estimate as it is completely accurate to the system's strength.

## Post Peak Recon Data, Satellite & Microwave Analysis

<u>Flight Level Data Analysis</u>

Post Lane's peak at around 0600 UTC, recon still found strong, although weaker flight level winds at 740-790hPa, in the north and south portions of the eyewall. The flight level winds that will be presented below are from the Atlantic Oceanographic & Meteorological Laboratory's Hurricane Research Division:

| TIME | Lat | Lon | Head | Track | GnSpd | TAS | GeoAl | Press | WndDr | WndSp | Tempr | Dewpt | D Val | RdAlt | MixR | VtWnd | SfcPr | ThetaE |
|---|---|---|---|---|---|---|---|---|---|---|---|---|---|---|---|---|---|---|
| HHMMSS | Deg N | Deg W | Deg | Deg | m/s | m/s | m | mb | Deg | m/s | C | C | m | m | g/m3 | m/s | mB | Deg |
| 060508 | 14.702 | 154.257 | 145.00 | 176.80 | 128.3 | 129.7 | 2267.0 | 746.8 | 71.70 | 72.00 | 13.41 | 14.00 | -232.3 | 2272.0 | 13.6600 | 2.10 | 970.0 | 353.60 |
| 060509 | 14.700 | 154.257 | 144.60 | 177.00 | 127.9 | 129.2 | 2267.0 | 746.6 | 71.70 | 72.20 | 13.35 | 13.90 | -234.0 | 2272.0 | 13.5300 | 0.90 | 969.9 | 353.20 |
| 060510 | 14.698 | 154.257 | 144.60 | 177.10 | 127.4 | 128.7 | 2267.0 | 746.5 | 71.80 | 72.00 | 13.21 | 13.80 | -235.1 | 2271.0 | 13.4100 | 0.00 | 969.9 | 352.60 |
| 060511 | 14.698 | 154.257 | 144.90 | 177.30 | 126.9 | 127.8 | 2266.0 | 746.4 | 71.30 | 72.80 | 13.13 | 13.70 | -237.9 | 2271.0 | 13.4000 | 0.30 | 969.6 | 352.50 |
| 060512 | 14.697 | 154.257 | 144.90 | 177.40 | 126.4 | 126.9 | 2265.0 | 746.2 | 71.10 | 72.70 | 13.03 | 13.70 | -240.7 | 2270.0 | 13.4100 | 1.10 | 969.4 | 352.40 |
| 060513 | 14.695 | 154.257 | 144.50 | 177.50 | 126.0 | 126.1 | 2265.0 | 746.2 | 70.70 | 73.10 | 12.91 | 13.60 | -240.7 | 2269.0 | 13.3200 | 1.20 | 969.5 | 352.00 |
| 060514 | 14.695 | 154.257 | 143.70 | 177.40 | 126.1 | 126.9 | 2263.0 | 746.2 | 71.20 | 73.50 | 12.83 | 13.60 | -242.3 | 2268.0 | 13.2600 | 0.90 | 969.4 | 351.70 |
| 060515 | 14.693 | 154.257 | 142.90 | 177.20 | 126.1 | 127.6 | 2261.0 | 745.6 | 71.60 | 73.00 | 12.76 | 13.50 | -250.5 | 2266.0 | 13.2400 | 1.90 | 968.5 | 351.70 |
| 060516 | 14.692 | 154.257 | 143.00 | 177.10 | 125.9 | 126.8 | 2260.0 | 745.4 | 71.20 | 72.10 | 13.10 | 13.50 | -254.5 | 2264.0 | 13.2200 | 4.00 | 967.7 | 352.10 |
| 060517 | 14.692 | 154.257 | 144.00 | 177.10 | 125.6 | 124.9 | 2259.0 | 745.3 | 69.70 | 72.10 | 13.40 | 13.50 | -256.5 | 2264.0 | 13.2300 | 4.70 | 967.2 | 352.50 |
| 060951 | 14.390 | 154.227 | 207.20 | 179.00 | 116.8 | 126.8 | 1800.0 | 784.3 | 278.20 | 69.40 | 14.62 | 17.40 | -308.2 | 1803.0 | 16.2100 | 6.30 | 965.1 | 357.80 |
| 060952 | 14.388 | 154.227 | 210.40 | 178.90 | 117.5 | 122.6 | 1793.0 | 785.2 | 280.30 | 64.10 | 15.01 | 17.40 | -305.1 | 1797.0 | 16.2000 | -1.10 | 965.3 | 358.10 |
| 060953 | 14.388 | 154.227 | 212.50 | 179.20 | 117.1 | 126.1 | 1784.0 | 786.3 | 276.80 | 63.80 | 14.89 | 17.40 | -303.4 | 1787.0 | 16.1400 | -0.60 | 965.7 | 357.60 |
| 060954 | 14.387 | 154.227 | 211.80 | 179.80 | 116.6 | 127.5 | 1774.0 | 787.7 | 275.60 | 64.60 | 15.01 | 17.40 | -298.5 | 1778.0 | 16.1100 | -1.50 | 966.3 | 357.50 |
| 060955 | 14.385 | 154.227 | 210.10 | 179.70 | 116.2 | 128.9 | 1769.0 | 788.2 | 274.30 | 65.90 | 14.95 | 17.40 | -298.8 | 1772.0 | 16.1100 | -2.20 | 966.4 | 357.30 |
| 060956 | 14.385 | 154.227 | 209.80 | 179.60 | 115.6 | 124.2 | 1765.0 | 788.9 | 276.90 | 62.30 | 15.60 | 17.40 | -295.6 | 1768.0 | 16.0800 | -0.30 | 966.4 | 358.00 |
| 060957 | 14.383 | 154.227 | 210.30 | 179.30 | 115.0 | 125.8 | 1764.0 | 789.6 | 276.00 | 65.80 | 15.62 | 17.40 | -289.5 | 1767.0 | 16.0600 | 2.20 | 967.1 | 357.90 |
| 060958 | 14.383 | 154.223 | 211.40 | 179.20 | 114.5 | 117.9 | 1765.0 | 790.0 | 280.80 | 58.80 | 16.23 | 17.40 | -284.0 | 1768.0 | 16.0600 | 1.30 | 967.3 | 358.60 |
| 060959 | 14.382 | 154.227 | 210.70 | 179.50 | 113.6 | 117.8 | 1765.0 | 790.1 | 279.80 | 57.60 | 16.08 | 17.40 | -282.7 | 1769.0 | 16.0700 | -2.00 | 967.6 | 358.40 |
| 061000 | 14.380 | 154.227 | 207.80 | 179.30 | 114.0 | 120.0 | 1764.0 | 790.2 | 279.50 | 63.50 | 15.47 | 17.40 | -282.8 | 1768.0 | 16.0800 | 0.80 | 968.0 | 357.70 |

*Figure 20 & 21: Flight level winds during 0605 UTC in the north portion of the eyewall, and flight level winds during 0609 UTC in the south portion of the eyewall.*

At 060514 UTC in the north portion of the eyewall, recon recorded a maximum wind of 73.5 meters per second, or 142.8kts. However, this is one second data. For winds to be countable as representative of a storm's 1-min intensity, a 10 second average once again must be taken. Averaging the values in the dataset achieves a 10 second average of 72.6 meters per second, or 141.1kts. Taking an average of the radar altitude values gets a mean height of 2268 meters. Using this height on the vertical wind profile gets a surface conversion of x0.97. Applying the conversion to the 141.1kt 10 second average gets a surface value of 136.9kts. Without considering anything outside of the flight level data and only vertical wind profile results, this would support a 135kt intensity around 2 hours after peak.

At 060951 UTC in the south portion of the eyewall, recon recorded a maximum wind of 69.4 meters per second, or 134.9kts. However, once again this is one second data. Averaging the values in the dataset gets a 10 second average of 63.6 meters per second, or 123.6kts. Taking an average of the radar altitude values gets a mean height of 1778 meters. Using this height on the vertical wind profile gets a surface conversion of x0.95. Applying the conversion to the 123.6kt 10 second average gets a surface value of 116.9kts.

Something that immediately stands out about the flight level data, is how asymmetrical the wind field is, plotted below:

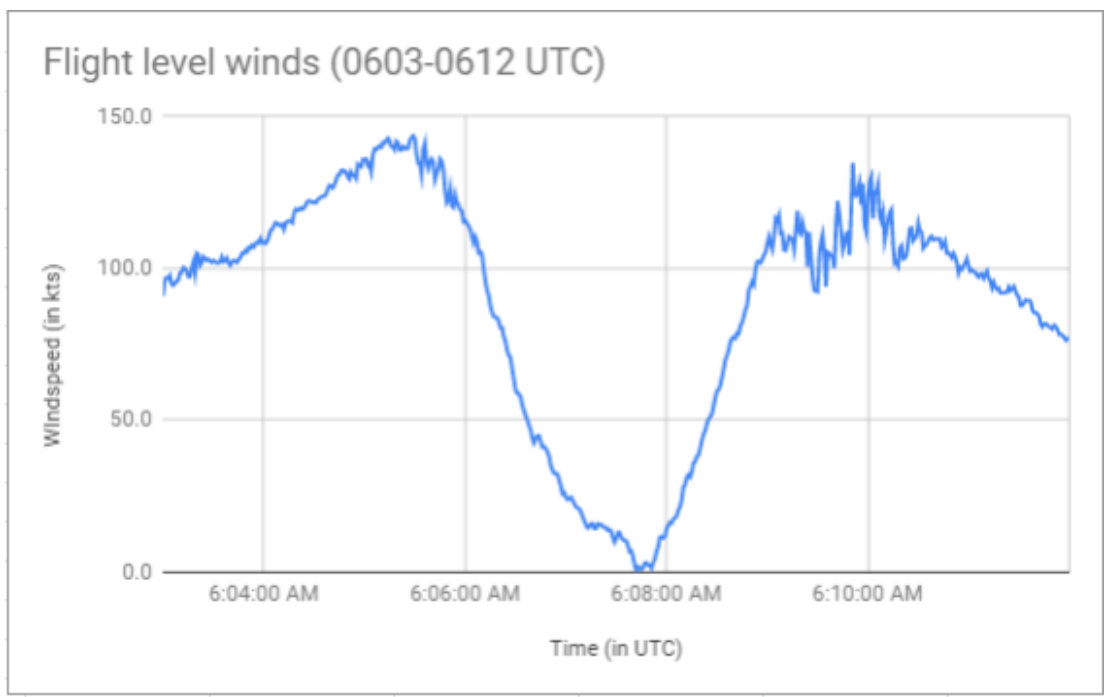

*Figure 22: Flight level wind plot shows how asymmetrical Lane's wind field was.*

This asymmetry can partially be attributed to Lane's movement speed and direction, as at 0600 UTC, it was moving at 8kts due west. This movement speed would be enough to cause some wind field asymmetry, especially in a weakening system like Lane.

Another issue that stands out is the data shown in the south section of the eyewall. It is extremely unstable, with winds jumping up and down several knots each second. The reason this occurred is due to extreme turbulence the aircraft experienced while in the south eyewall, shown on static aircraft pressure plotted below:

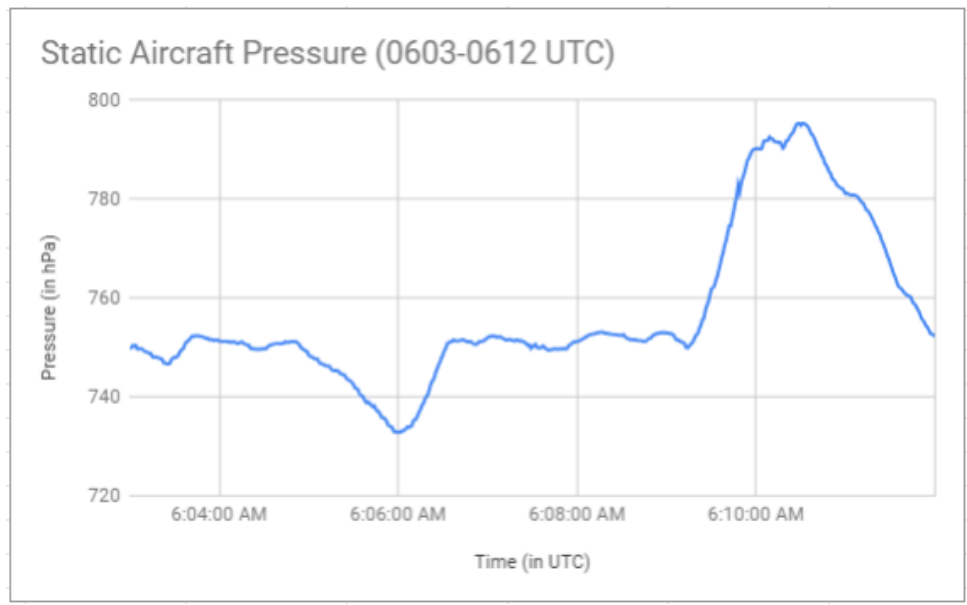

*Figure 23: Static aircraft pressure plotted, showing turbulence in the south section of the eyewall.*

At around 0609 UTC, while recon experienced the unstable flight level readouts, the plane descended and ascended by 3hPa in less than a minute. This is a hallmark of extreme turbulence, along with the various other rapid descents and ascents recorded by the aircraft within this time span. This most likely affected the reliability of the anemometer that records flight level wind, and as a result caused the noisy and spiky data. This turbulence was also confirmed by crew onboard the aircraft, shown on the next page:

*Figures 24 & 25: Notes from the recon mission.*

In the southern eyewall, where these recordings were located, it was so turbulent that the plane consistently recorded 2.5+ Gs, and the pilots eventually had to abort as they could not find an area to penetrate that would not mess up data due to high turbulence. This further supports that the data could be off for the true intensity of the south section of the eyewall, as vibrations from the extreme turbulence, also clear in data would cause issues. This will be referred to later in the analysis, with a further investigation into Lane's wind symmetry.

Similarly, to the peak analysis, the surface conversions and their 6.76% uncertainty range for each eyewall section will be plotted in a table below to conclude the flight level data analysis:

| Confidence Rates & Conversion Results | "Low" 95% | Initial Conversion | "High" 95% |
|---|---|---|---|
| North Section | 127.6kts | 136.9kts | 146.2kts |
| South Section | 109.0kts | 116.9kts | 124.8kts |

*Figure 26: A chart showing low- and high-end confidence rates for each recorded section of the eyewall, resulting from a less than optimal amount of datapoints.*

SFMR Data Analysis

Post Lane's peak around 0600 UTC, SFMR began to record extreme values far stronger than maximum flight level winds what they suggested at surface again, albeit slightly weaker this time. The data that will be displayed below is once again from the Atlantic Oceanic Meteorological Laboratory's Hurricane Reanalysis Division:

```
Time (in UTC)   RR     WS(m/s) | kts)    TB1     TB2     TB3     TB4     TB5     TB6
6:05:26 AM      7.3    75.9     147.5    208.3   215.6   219.5   225.3   231.8   235.3
6:05:27 AM      5.8    76.3     148.3    208.4   215.6   219.5   225.3   231.8   234.3
6:05:28 AM      4.1    76.6     148.9    208.4   217.3   220.2   225.3   231.8   234.3
6:05:29 AM      2.7    76.8     149.3    208.4   217.3   220.2   225.3   229.2   234.3
6:05:30 AM      1.9    76.7     149.1    210     217.3   220.2   225.3   229.2   231.8
6:05:31 AM      0.9    76.7     149.1    210     215.8   218.5   225.3   229.2   231.8
6:05:32 AM      0.2    76.6     148.9    210     215.8   218.5   222.8   226.2   231.8
6:05:33 AM      0      76.3     148.3    208.3   215.8   218.5   222.8   226.2   228.4
6:05:34 AM      0      75.9     147.5    208.3   215.3   216.9   222.8   226.2   228.4
6:05:35 AM      0      75.6     147.0    208.3   215.3   216.9   220.9   224.8   228.4

6:08:50 AM      31.1   67.3     130.8    201.1   209.8   215.1   222.1   231     233.9
6:08:51 AM      32.9   67.2     130.6    203.5   209.8   215.1   222.1   231     240
6:08:52 AM      33.8   67.4     131.0    203.5   212.7   217.2   222.1   231     240
6:08:53 AM      34.7   67.5     131.2    203.5   212.7   217.2   224.9   235.1   240
6:08:54 AM      35.8   67.4     131.0    203.4   212.7   217.2   224.9   235.1   241.8
6:08:55 AM      36.7   67.3     130.8    203.4   212.1   216.9   224.9   235.1   241.8
6:08:56 AM      36.8   67.3     130.8    203.4   212.1   216.9   223.9   232.8   241.8
6:08:57 AM      37.3   67.2     130.6    202.9   212.1   216.9   223.9   232.8   238.3
6:08:58 AM      37.5   67.1     130.4    202.9   210.9   215.6   223.9   232.8   238.3
6:08:59 AM      37.5   67       130.2    202.9   210.9   215.6   222.6   231.8   238.3
```

*Figure 27 & 28: SFMR winds during 0605 UTC in the north portion of the eyewall, and SFMR winds during 0608 UTC in the southern portion of the eyewall.*

At 060529 UTC in the north section of the eyewall, SFMR estimated 149.3kt winds for one second, which still marks a clear discrepancy between flight level winds and their conversions, and SFMR winds. A 10 second average of the available data gets a 10 second wind of 148.4kts, which is still be remarkably higher than flight level winds and their conversions.

Something that is noteworthy about the data, is that there seems to be almost no rain affecting it, with a maximum rain rate of 2.7 mm/hr. during the peak value, lowering the risk of direct contamination by higher rain rates. Without further inspection, these values would be considered accurate to Lane's intensity. An additional note is that these readings seem similar to the readings found during peak mission, in the sense that extremely high SFMRs relative to flight level winds are recorded, with low rain rates that would make them seem accurate at first glance.

At 060853 UTC in the east section of the eyewall, SFMR estimated 131.2kt winds for one second, which is still higher in comparison to the flight level winds recorded at the time. A 10 second average of the available data gets a 10 second wind of 130.7kts. This data has a high rain rate at peak winds of 34.7 mm/hr. This rain rate is potentially enough to cause issues with the recorded data, thus making it less reliable without further review. SFMR compared to flight level winds are shown on the next page:

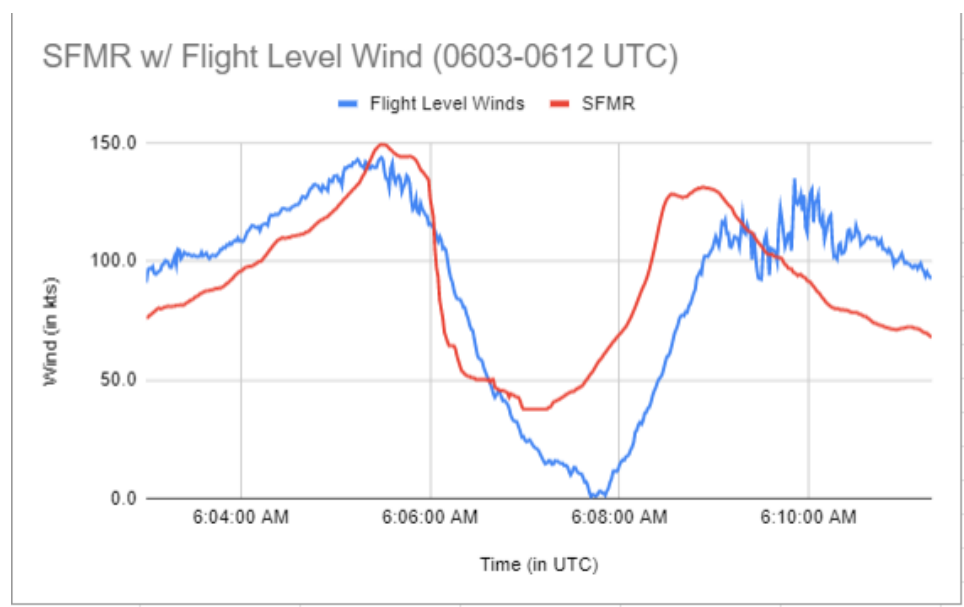

*Figure 29: Flight level winds plotted alongside SFMR winds.*

Something that is easily noticeable is lack of wind symmetry once again, this time shown by both flight level winds and SFMR. However, this time, instead of it being entirely instrument error, it's likely also caused the weakening system combined with motion vector and speed.

Something else that can be inferred is that the rather extreme stadium effect shown by recon observations and data, is not as apparent, along with no observations being made on any VDMs talking about it. However, it is still visible on the data with the surface radius of maximum winds peaking before the flight level radius of maximum winds. It is also still apparent with the lower rain rates, as explained earlier in the analysis.

Plotting SFMR with temperature brightness data and rain rates still reveals something more similar to what was recorded with the peak passes, shown on the next page:

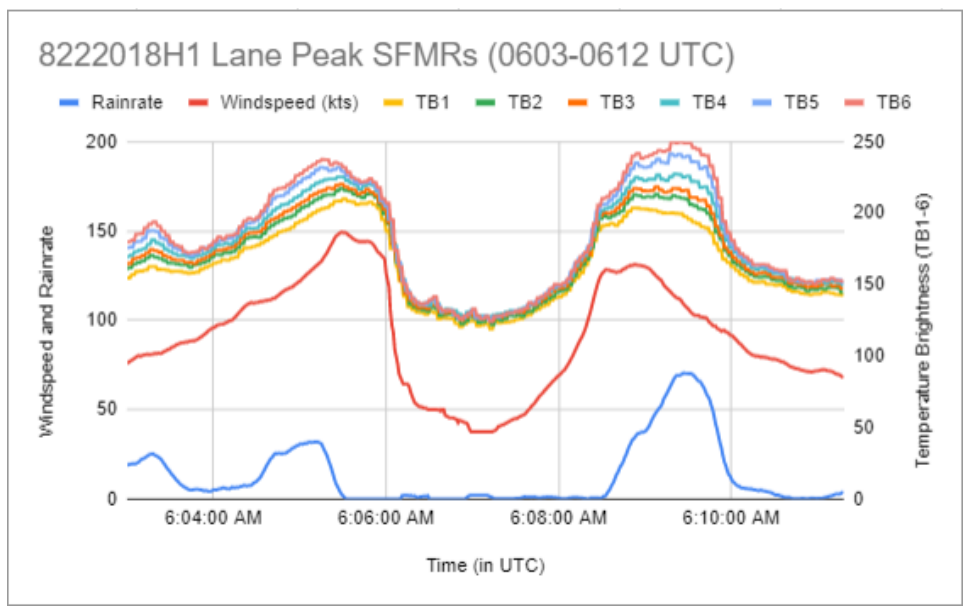

*Figure 30: SFMR winds plotted with rain rates and temperature brightness signals.*

While there still is a relationship between spiking rain rates and temperature brightness when recon recorded 140kt+ SFMRs, it is admittedly not as obvious as before, as there are far lower rain rates to affect temperature brightness values than at peak. However, it is recommended to still take these signals and their resultant wind estimates with caution, as it is still likely that it is inflated to a degree. After recon crossed the eye, there was a far more notable rain rate spike that was more closely correlated with temperature brightness signal spikes, which suggests there could have been some inflation there. However, as rain rate continues to spike, eventually temperature brightness signals almost hard cap along with wind estimates, which could once again be the algorithm accounting for potential inflation from rain. Regardless, the wind estimate resulting from the south quadrant is suspect taking into account what the plot is showing, along with some potential recording issues resulting from the extreme turbulence the aircraft went through.

Research revealing a high bias in SFMR estimation above 120-130kts also still raises some more questions over if this data, especially the data recording around or above 150kts, is valid at all, and how much the algorithm accounted for the possible temperature brightness inflation from shear forcing. Shear forcing in this scenario is when a tropical cyclone gets powerful enough, around the same threshold that the bias seems to start occurring at (120-130kts), the ocean surface does not exactly seem to get rougher, but actually smoother instead. However, what would occur in place is that more sea spray would be sheared off to achieve a smoother ocean surface. This could cause a bigger problem when it comes to SFMR, as the algorithm is heavily dependent on temperature brightness values that result from sea spray, and an inflation of such would inflate wind estimates past what they should be. Additionally, this shear forcing causing lower roughness would inherently cause lower friction, which as a result, the boundary layer would go down, as it is a function of lower-level friction, resulting in the lower boundary layer shown by dropsondes in the VWP section of the analysis. However, this is admittedly out of the scope of this section of the analysis.

As a result of all aforementioned reasons, SFMR data from post peak Lane should also not be taken at face value, and should simply be used alongside other methods.

Dropsonde Analysis

During Lane's peak, there unfortunately were no useful eyewall dropsondes to aid in estimating intensity using WL150 or MBL500. However, during post peak, there was one reliable eyewall dropsonde that was also dropped during a flight level maximum in the north section of the eyewall. The dropsonde and its gusts, WL150 and MBL500 is shown on the next page:

| Significant Wind Levels | | |
|---|---|---|
| Level | Wind Direction | Wind Speed |
| 944mb (Surface) | 310° (from the NW) | 140 knots (161 mph) |
| 939mb | 315° (from the NW) | 145 knots (167 mph) |
| 936mb | 325° (from the NW) | 171 knots (197 mph) |
| 935mb | 320° (from the NW) | 157 knots (181 mph) |
| 934mb | 325° (from the NW) | 169 knots (194 mph) |
| 929mb | 340° (from the NNW) | 151 knots (174 mph) |
| 928mb | 340° (from the NNW) | 147 knots (169 mph) |
| 927mb | 345° (from the NNW) | 153 knots (176 mph) |
| 913mb | 350° (from the N) | 151 knots (174 mph) |
| 901mb | 0° (from the N) | **173 knots (199 mph)** |
| 898mb | 0° (from the N) | 168 knots (193 mph) |
| 893mb | 5° (from the N) | 150 knots (173 mph) |
| 887mb | 10° (from the N) | 141 knots (162 mph) |
| 885mb | 10° (from the N) | 143 knots (165 mph) |
| 881mb | 10° (from the N) | 157 knots (181 mph) |
| 877mb | 10° (from the N) | 153 knots (176 mph) |
| 875mb | 20° (from the NNE) | 144 knots (166 mph) |
| 873mb | 20° (from the NNE) | 151 knots (174 mph) |
| 865mb | 25° (from the NNE) | 162 knots (186 mph) |
| 861mb | 30° (from the NNE) | 151 knots (174 mph) |
| 850mb | 35° (from the NE) | 140 knots (161 mph) |
| 834mb | 35° (from the NE) | 128 knots (147 mph) |
| 801mb | 50° (from the NE) | 151 knots (174 mph) |
| 798mb | 60° (from the ENE) | 133 knots (153 mph) |
| 791mb | 55° (from the NE) | 135 knots (155 mph) |
| 789mb | 60° (from the ENE) | 120 knots (138 mph) |
| 733mb | 70° (from the ENE) | 114 knots (131 mph) |

**Mean Boundary Level Wind (mean wind in the lowest 500 geopotential meters of the sounding):**
- **Wind Direction:** 345° (from the NNW)
- **Wind Speed:** 149 knots (171 mph)

**Deep Layer Mean Wind (average wind over the depth of the sounding):**
- **Wind Direction:** 30° (from the NNE)
- **Wind Speed:** 120 knots (138 mph)
- **Depth of Sounding:** From 733mb to 944mb

**Average Wind Over Lowest Available 150 geopotential meters (gpm) of the sounding:**
- **Lowest 150m:** 147 gpm - -3 gpm (482 geo. feet - -10 geo. feet)
- **Wind Direction:** 325° (from the NW)
- **Wind Speed:** 152 knots (175 mph)

*Figure 31 & 32: Dropsonde gusts, MBL500 and WL150 shown above*

The dropsonde's MBL500 will be multiplied by x0.8 due to research (Franklin et al. 2002) revealing a mean 10-meter wind to meal boundary layer wind ratio of 0.8. Multiplying the 149kt MBL500 by x0.8 gets a surface wind of 119.2kts. However, with the MBL500 comes problems. This conversion is admittedly conservative since not many tropical cyclones have a 500 GPM tall boundary level. This allows for such data to be affected by non-boundary level winds, deflating the values achieved from the average. The MBL conversion also has a weakness in which it treats every storm alike. The conversion assumes that all storms will have the same conversion of x0.8 when that is likely not the case. The multiplier can be altered, but unfortunately there is not enough eyewall dropsondes like this accurate to the storm's intensity to determine a new MBL500 multiplier from.

The dropsonde's WL150 will be multiplied by a x0.85 multiplier derived from an average of its lowest 150-meter heights of 147 meters and –3 meters, then locating the average on the fit below:

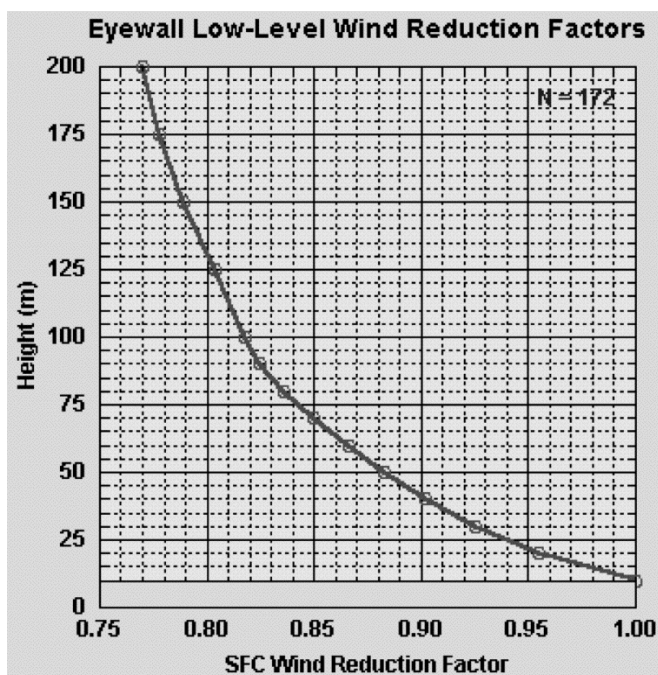

*Figure 33: Height to SFC wind reduction factor chart used to achieve the multiplier for WL150*

Multiplying the 152kt by x0.85 yields a surface wind of 129.2kts. This seems like it would be accurate. However, there are further issues with a device like dropsondes with detecting surface winds accurate to a storm's intensity. Dropsondes do not sample in a swath, but only sample what gusts they are impacted by. As a result, under sampling is a larger issue than what it seems at first glance. Furthermore, another reason for the dropsonde likely not being extremely accurate, is that Lane still had a stadium effect at that point. The dropsonde could have easily fallen out of the radius of maximum winds, as dropsondes are not able to be dropped in a way where it can sample extreme stadium effects perfectly. As a result of it probably not getting the surface radius of maximum winds, a larger range of uncertainty can also be applied, and this will be taken into account later in the analysis

Aside from the eyewall dropsonde, there was a dropsonde dropped in the eye of Lane to estimate its central pressure. The dropsonde recorded a pressure of 932hPa at surface. Furthermore, the 932hPa pressure was accompanied by an 8kt gust. An 8kt gust would signify that this was very close to the exact center of the cyclone, and as a result, the pressure needs no adjustment. The dropsonde is shown below:

| Significant Wind Levels | | |
|---|---|---|
| Level | Wind Direction | Wind Speed |
| 932mb (Surface) | 235° (from the SW) | 8 knots (9 mph) |
| 927mb | 240° (from the WSW) | **13 knots (15 mph)** |
| 900mb | 260° (from the W) | 9 knots (10 mph) |
| 850mb | 275° (from the W) | 12 knots (14 mph) |
| 750mb | 260° (from the W) | 1 knots (1 mph) |

*Figure 34: The dropsonde that measured a 932hPa pressure*

Recon Radar & Recon Data

Recon velocity radar data can still help make sense of the recon data previously analyzed. The velocity radar image at 2500 meters (around the height recon flew at) is shown on the next page:

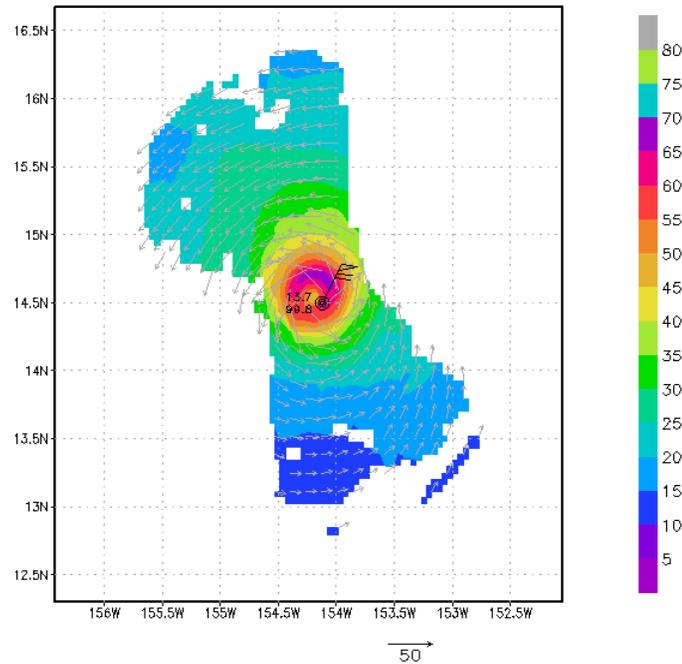

*Figure 35: Velocity radar map of Lane at 2500 meters*

The radar data coincidentally shows the same thing as flight level winds and SFMR, wind asymmetry. This would help verify the weaker winds in the south portion of the eyewall than north portion of the eyewall resulting from movement speed, vector, and the storm weakening as well. Overall, this velocity radar data would help prove that the asymmetry was not mainly a result of turbulence or anything else that would cause recording issues.

Comparing Satellite Imagery

One thing that can be used to estimate Lane's peak intensity outside of purely recon methods, is backward extrapolation of intensity from a recon observed time like 0600 UTC, to 0300 UTC. The results of this section are automatically bound to some error as it is not as conclusive. However, it can still server as a valid backing point for estimating peak intensity. Two images of Lane at peak, then post peak will be shown on the next page:

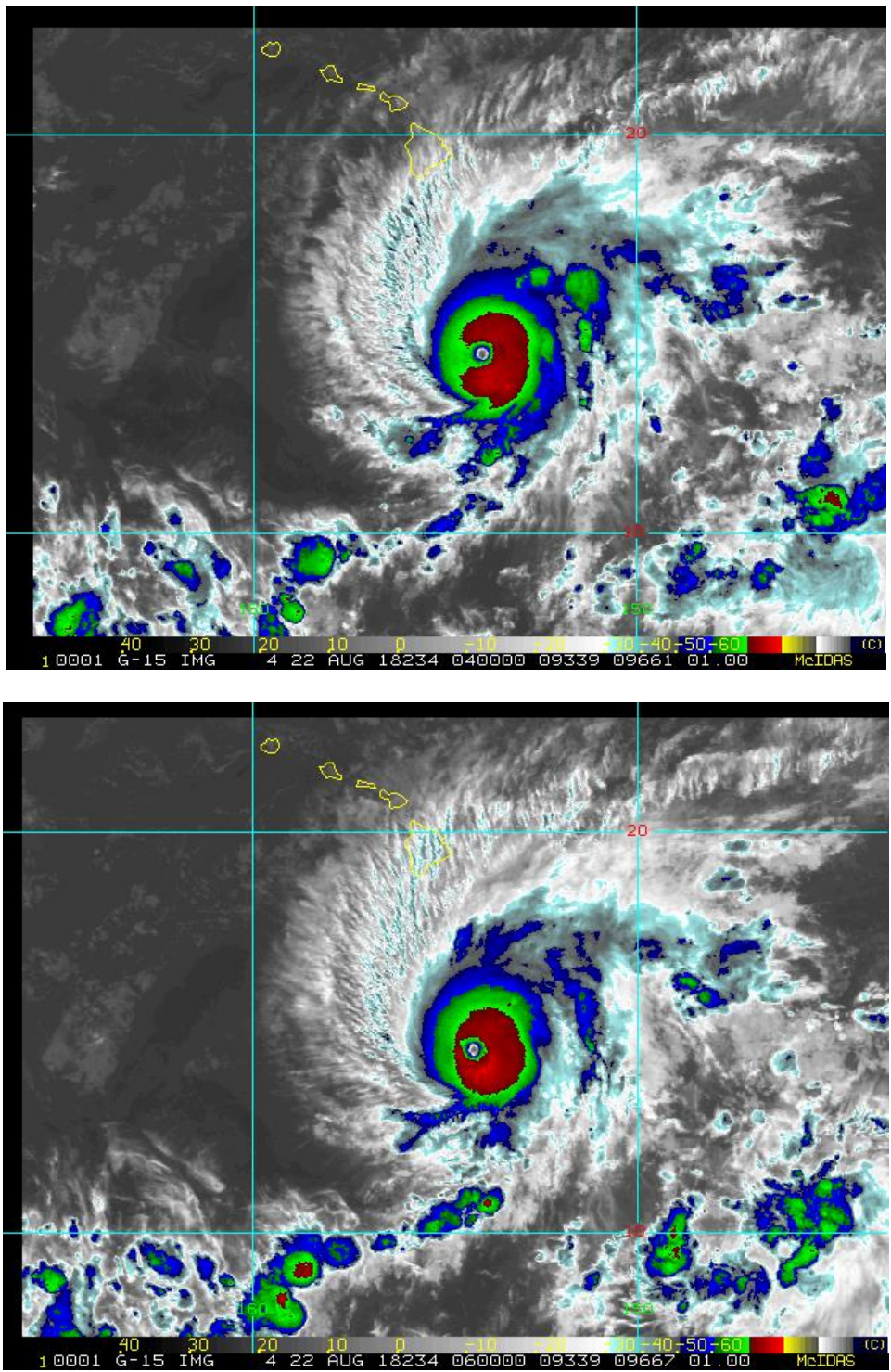

*Figure 35 & 36: In order, images of Lane at peak and post peak*

In the images above, a difference that would deem post peak slightly weaker than peak, is a cooling eye. As systems maintain their intensity or strengthen maybe, their eye does not cool. This is due to the fact that strong convective bursts resulting from the maintaining or strengthening are allowing for subsidence from the bursts to maintain the dry central column, or keep drying it. As a system weakens and begins to collapse, convection would not fare well at keeping the central column dry. As the storm weakens, updraft should presumably weaken as well, which is what allows for tropical cyclones to get their significant convection anyways. Regardless, if Lane's eye has cooled and gotten moister, it is likely that it weakened from its previous state.

Another thing that would deem it weaker, is the lack of a more united lift evident in the images. In the first image at peak, while Lane struggled to lift the west side of the CDO due to westerly shear, it is still evident that the CDO had far less variance in temperature, which would be typical of an intense tropical cyclone, since as they mature, they enter a mode where convection evolves from a more super rotational convective burst pattern to a fuller, smoother pattern. Lane begins to lack that at 0600 UTC in the eastern part of the CDO, in the south, where a part of the CDO is visibly cooler than the rest, signaling a more uneven lift and that it has likely weakened. It can also be inferred that it lost neutral buoyancy, as it showed that during peak. Neutral buoyancy is when a tropical cyclone's CDO temp doesn't seem to vary much, if at all, and is usually the hallmark of intense, stable tropical cyclones.

A slight expansion to the east is also visible, which along with all other structural signs elaborated on previously, could indicate the start of an eyewall replacement cycle, which would weaken a tropical cyclone as the pressure gradient broadens, inner eyewall dissipates, and larger outer eyewall takes over.

Finally, the storm's stadium effect is more oblong compared to its previous symmetrical and well-defined state. This would also indicate lack of convective organization, which is typical of a weakening tropical cyclone. Overall, from all of the aforementioned circumstances, it is fair to assume that Lane weakened ~5kts from its peak intensity, which will be fully defined later in the analysis.

## Conclusion

Lane's post peak intensity will be calculated in a way that's similar to that of the peak intensity. The vertical wind profile is still more certain than the WL150 and SFMRs from analysis of their sections, so as a result it will be double weighted in the geometric mean calculation and neither the WL150 or SFMR will be weighted over one another. The calculation will be shown below:

$$(136.9 \cdot 136.9 \cdot 129.2 \cdot 148.4)^{\frac{1}{4}}$$

*Figure 37: The double weighted surface winds from flight level conversions, converted WL150, and peak 10 second average of SFMR winds in a geometric mean equation.*

The results of this calculation yield a 137.7kt mean, which rounds best to 140kts. As a result, 140kts along with the pressure of 932hPa from earlier will be the derived intensity for 0600 UTC, and this would help prove peak intensity, as the author inferred ~5kts of weakening from that point based on satellite trends.

## Final Estimate

Based on a more holistic approach of all evidence pertaining to what Lane's intensity could be, the author has concluded a **145kt/926hPa** intensity at Lane's peak around 0400 UTC. Due to some uncertainty still existing as a result of the large disagreements between objective measurements, the author has also adopted the standard reconnaissance range of uncertainty, which is +/- 5kts.

## Acknowledgements

The author of this analysis would like to thank some people for their tips and helpful advice. The author would like to thank the entire "Analysis Gulag" group chat for peer reviewing this analysis firstly. Secondly, he would like to thank wer for providing the range of uncertainty equation used in the flight level analysis sections, thirdly he would like to thank mck for reviewing the vertical wind profile analysis and giving very helpful tips on how to replicate Franklin's study, and KeviShader for providing a brief look through of the analysis as well. Finally, the author would like to thank the scientists and pilots on the flights into Lane that ultimately got the data that was used in this comprehensive analysis. Without them, it would not have been possible at all.